\documentclass[amsmath,amssymb,prl,twocolumn,unsortedaddress]{revtex4-2}

\usepackage{graphicx,natbib}
\usepackage{dcolumn}
\usepackage{bm}
\usepackage{hyperref}
\usepackage{xcolor}
\usepackage{soul}

\newcommand{\vecB}{\textbf{B}}
\newcommand{\vecE}{\textbf{E}}
\newcommand{\vecv}{\textbf{v}}
\newcommand{\vecu}{\textbf{u}}
\newcommand{\vecJ}{\textbf{J}}
\newcommand{\matP}{\textbf{P}}
\newcommand{\matPi}{\mathbf{\Pi}}
\newcommand{\matD}{\textbf{D}}
\newcommand{\matT}{\textbf{T}}
\newcommand{\JdotE}{\vecJ\cdot\vecE}
\newcommand{\JidotE}{\vecJ_i\cdot\vecE}
\newcommand{\JedotE}{\vecJ_e\cdot\vecE}
\newcommand{\PiD}{\matPi : \matD}
\newcommand{\PieDe}{\matPi_e : \matD_e}
\newcommand{\PiiDi}{\matPi_i : \matD_i}
\newcommand{\ptheta}{p\theta}
\newcommand{\pthetae}{p_e\theta_e}
\newcommand{\pthetai}{p_i\theta_i}

\newcommand{\EEM}{E^\mathrm{EM}}
\newcommand{\Eth}{E^\mathrm{th}}
\newcommand{\Ef}{E^\mathrm{f}}
\newcommand{\Epar}{E_{\|}}
\DeclareMathAlphabet\mathbfcal{OMS}{cmsy}{b}{n}

\begin{document}

\title{Kinetic heating by Alfv\'en waves in magnetic shears}

\author{Fabio Bacchini}
 \email{fabio.bacchini@kuleuven.be}
 \affiliation{Centre for mathematical Plasma Astrophysics, Department of Mathematics, Katholieke Universiteit Leuven, Celestijnenlaan 200B, B-3001 Leuven, Belgium}
\author{Francesco Pucci}\email{francesco.pucci@kuleuven.be}
 \affiliation{Istituto per la Scienza e Tecnologia dei Plasmi, Consiglio Nazionale delle Ricerche (ISTP-CNR), Via Amendola 122/D, 70126 Bari, Italy}
 \affiliation{Centre for mathematical Plasma Astrophysics, Department of Mathematics, Katholieke Universiteit Leuven, Celestijnenlaan 200B, B-3001 Leuven, Belgium}
\author{Francesco Malara}
 \affiliation{Dipartimento di Fisica, Universit\`{a} della Calabria, 87036 Rende (CS), Italy}
\author{Giovanni Lapenta}
 \affiliation{Centre for mathematical Plasma Astrophysics, Department of Mathematics, Katholieke Universiteit Leuven, Celestijnenlaan 200B, B-3001 Leuven, Belgium}

\begin{abstract}
With first-principles kinetic simulations, we show that a large-scale Alfv\'en wave (AW) propagating in an inhomogeneous background decays into kinetic Alfv\'en waves (KAWs), triggering ion and electron energization. We demonstrate that the two species can access unequal amounts of the initial AW energy, experiencing differential heating. During the decay process, the electric field carried by KAWs produces non-Maxwellian features in the particle VDFs, in accordance with space observations.
The process we present solely requires the interaction of a large-scale AW with a magnetic shear and may be relevant for several astrophysical and laboratory plasmas.
\end{abstract}


\maketitle

{\it Introduction.} Alfv\'en waves (AWs) are low-frequency electromagnetic fluctuations that can originate and propagate in magnetized plasmas. AWs have been predicted \cite{alfven1942existence} and later observed in the solar wind and in many other space and laboratory plasmas \cite{belcher1971large,smith1995ulysses,chaston1999fast,gekelman1999review,cirtain2007evidence,tomczyk2007alfven}; they carry equal amounts of kinetic and magnetic energy, and have been invoked to significantly contribute to plasma energization in various astrophysical environments \cite{hollweg1981alfven,brunetti2004alfvenic,keiling2009alfven,antolin2010role,petrosian2012stochastic,lazarian2012turbulence,grant2018alfven,comisso2018particle}. In order to efficiently convert AW energy into particle energy, the original fluctuation must decay from macroscopic (fluid) scales to smaller (kinetic) scales. 

It is widely accepted that this decay can be efficiently promoted by the interaction of counter-propagating AWs~\cite{iroshnikov1964turbulence,kraichnan1965inertial,ng1996interaction,matthaeus1999coronal,howes2013alfven,pezzi2017revisiting}, leading to turbulence~\cite{politano1995model,howes2008kinetic,wan2012intermittent,chandran2015intermittency,franci2015high,servidio2015kinetic,matthaeus2021turbulence}. However, when the magnetic field through which an AW propagates is inhomogeneous, the wave decay can be directly mediated by the interaction of the initial large-scale fluctuation with the background medium~\cite{hollweg1987resonance,malara1996formation,petkaki1998topological,tsiklauri2002three,malara2003alfven,pucci2014evolution,kumar2020frb}. Contrary to the case of turbulence, the effects of this decay route on plasma heating at kinetic scales have remained as of yet largely unexplored~\cite{tsiklauri2011particle,tsiklauri2012three,chen2020}.

In this work we present the results of a numerical experiment in which the decay of AWs into kinetic AWs (KAWs) in a magnetic shear
is studied self-consistently over a wide range of length scales, from macroscopic (fluid) to microscopic (electron) scales.
Our initial configuration is such that the AW-to-KAW transition, promoted by an inhomogeneous background, results from the so-called phase-mixing process~\cite{heyvaerts1983coronal,de1999phase}.
We show that phase mixing leads to differential heating for ions and electrons. This differential heating is justified via a simple argument on how the two species can access the kinetic and magnetic energy carried by AWs.

\begin{figure*}
\centering
\includegraphics[width=1\textwidth, trim={1mm 72mm 33mm 0mm}, clip]{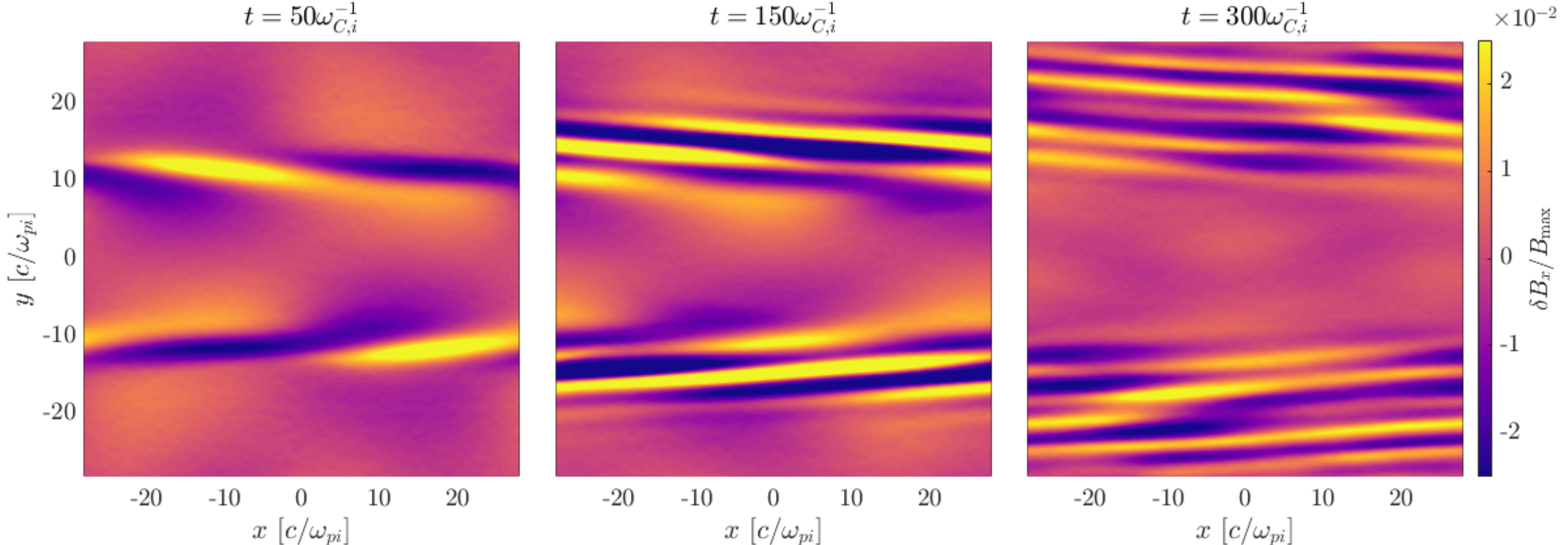}
\caption{Fluctuations in $B_x$ at $t=50,150,300\omega_{C,i}^{-1}$ (from left to right), identifying the KAWs produced inside, and drifting away from, the shear.}
\label{fig:dBx}
\end{figure*}

{\it Model.} We employ the semi-implicit Particle-in-Cell code \textsc{iPic3D}~\cite{markidis2010} to solve the Vlasov-Maxwell system of equations for an ion-electron plasma in a periodic two-dimensional $(x,y)$ Cartesian geometry. 
An inhomogeneous background magnetic field is initialized along $x$ with a sheared profile along $y$,
\begin{equation}
  B_{x,0}(y) = B_\mathrm{min} + \frac{B_\mathrm{max}-B_\mathrm{min}}{1+[y/(hL)]^r} + \alpha\left(\frac{y}{L/2}-1\right)^2,
\end{equation}
where $h=0.2, r=10$. The field strength decreases through the shear from $B_\mathrm{max}$ at $y=0$ to $B_\mathrm{min}=B_\mathrm{max}/2$ at $y=\pm L/2$. A small correction $\alpha$ is added to ensure that $dB_{x,0}/dy=0$ at the domain borders~\cite{vasconez2015kinetic}. The spatial profile of the number density $n=n_e=n_i$ is obtained by imposing pressure equilibrium, $B_{x,0}^2(y)/(8\pi) + 2n(y)kT = \mathrm{constant}$, and by considering an equal and uniform ion/electron temperature $T_i=T_e=T$.
The thermal-to-magnetic pressure ratio is $\beta_i = \beta_e = n kT/(B_\mathrm{max}^2/8\pi)=0.5$ at $y=0$. At $t=0$, a large-scale AW of wavelength $2\pi/L$ is superimposed to the background magnetic field as a perturbation in the out-of-plane direction  $\hat{{\bf z}}$:
\begin{equation}
    \delta B_z(x) = a\cos\left(\frac{2\pi}{L}x\right),\;\;
    \delta v_z(x,y) =- \frac{\delta B_z(x)}{B_{x,0}(y)}c_A(y),
\end{equation}
where $c_A(y)=B_{x,0}(y)/\sqrt{4\pi(m_i+m_e) n(y)}$ is the Alfv\'en speed, and $a=B_\mathrm{max}/10$. 
The particle velocity distribution functions (VDFs) are initialized as drifting Maxwellians, imposing that the total current $\vecJ = \vecJ_e+\vecJ_i=c\nabla \times \vecB/(4\pi)$ and that the center-of-mass velocity $\vecu=\delta v_z \hat{{\bf z}}$. The initial electric field is set according to Ohm's law
\begin{equation}
  \vecE = -\frac{1}{c}(\vecu\times\vecB) + \frac{1}{en_ec}(\vecJ\times\vecB) - \frac{1}{e n_e}\nabla\cdot\matP_e + \mathbfcal{I}_e,
  \label{eq:ohmslaw}
\end{equation}
where $\matP_e$ is the electron pressure tensor, and electron inertia terms $\mathbfcal{I}_e$ are neglected. The numerical domain has size $L_x=L_y=L=56 c/\omega_{pi}$, where $c$ is the speed of light and $\omega_{pi}$ is the ion plasma frequency. In order to limit computational costs while including electron-scale physics, we choose a reduced mass ratio $m_i/m_e=25$. The computational grid has $N_{x}\times N_{y}=280^2$ cells, corresponding to a spatial resolution $\Delta x=\Delta y = c/\omega_{pe}$. The time step is set to $\Delta t=0.25\omega_{C,e}^{-1}$, where $\omega_{C,e}$ is the electron cyclotron frequency at $y=0$. We employ $64000$ particles per cell to reduce numerical noise. The Alfv\'en-to-light speed ratio and the electron plasma-to-cyclotron frequency ratio at $y=0$ are $c_A/c=0.02$ and $\omega_{pe}/\omega_{C,e}=10$. Because our background equilibrium is not an exact solution of the Vlasov-Maxwell equations, we observe mild oscillations appearing in the system's energy integrals over time. However, we have verified that no spurious energization arises in our results due to our choice of initial conditions (see Supplemental Material),

{\it Results.} The simulation is run until $t=300\omega_{C,i}^{-1} = 5.3 t_A$, where $t_A=L/c_A(y=0)$ is the Alfv\'en crossing time. The initial perturbation propagates at the local Alfv\'{e}n speed, which is larger at the center of the domain than at the sides, leading to phase mixing inside the shear region. This causes small-scale fluctuations to form inside the shear, where the original wave vector $k_x$ bends, developing a $k_y$ component that grows linearly in time. Previous works have shown that when these fluctuations reach the Larmor-radius scale (i.e.\ $k_y r_{L,i}\sim1$) in such a configuration, the Alfv\'en mode is converted into a kinetic Alfv\'en mode and KAWs are produced inside the shear region~\cite{vasconez2015kinetic,pucci2016alfven,valentini2017transition,maiorano2020kinetic}. These KAWs have nonnegligible group speed along $y$ and drift towards the edge of the domain. Figure \ref{fig:dBx} shows subsequent phases of KAW evolution and propagation, in the form of parallel magnetic field fluctuations $\delta B_x$.

\begin{figure}
\centering
\includegraphics[width=1\columnwidth, trim={1mm 68mm 160mm 0mm}, clip]{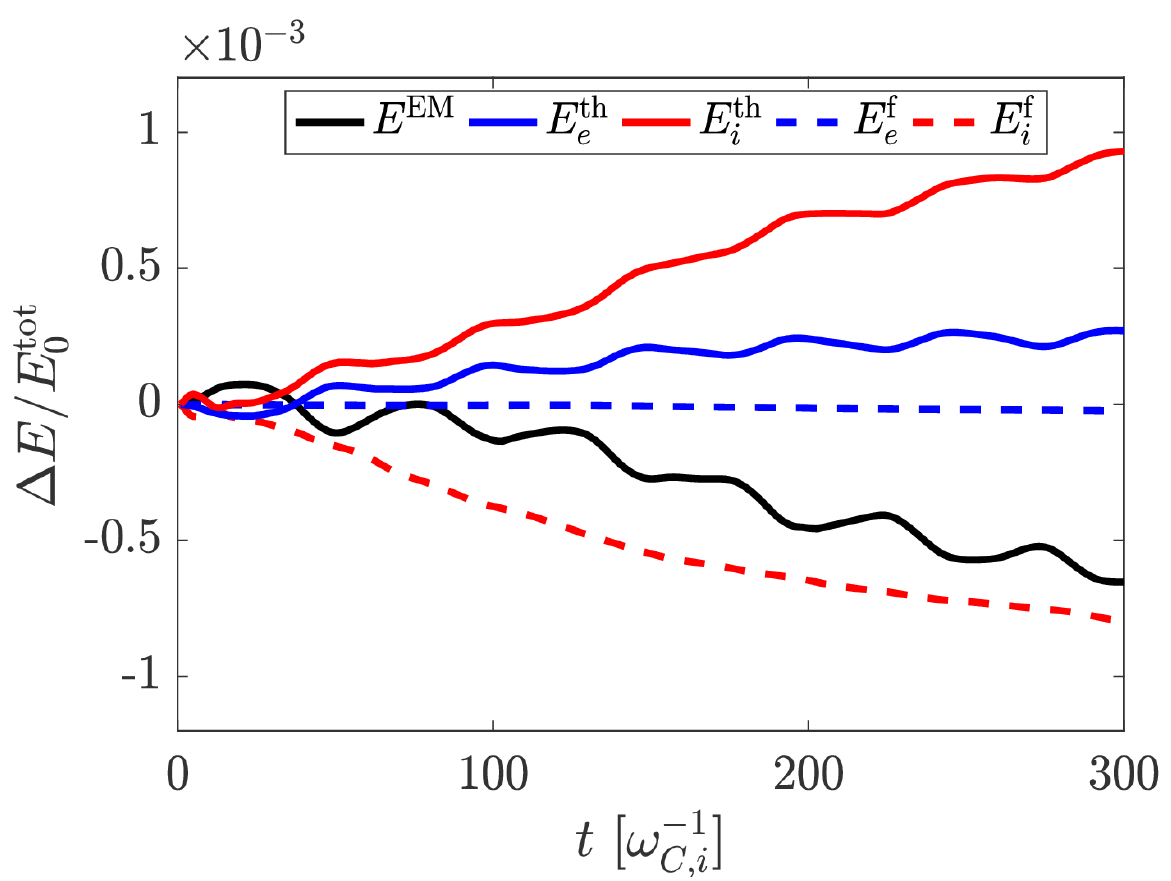}
\caption{Time variation of total electromagnetic energy, thermal energy, and bulk energy for electrons (subscript $e$) and ions (subscript $i$).}
\label{fig:dE}
\end{figure}

\begin{figure*}[t]
\centering
\includegraphics[width=1\textwidth, trim={0mm 15mm 0mm 0mm}, clip]{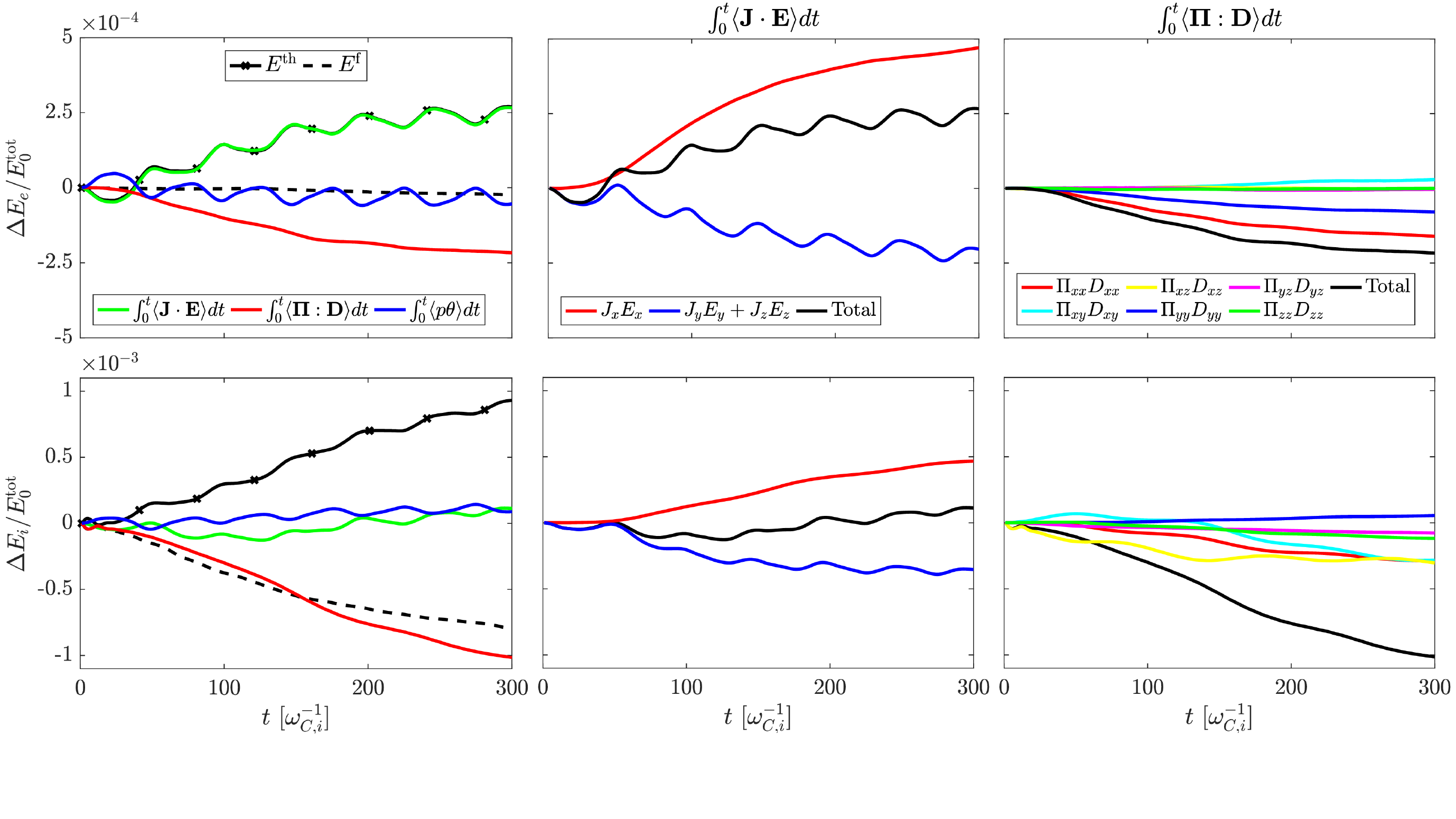}
\caption{Contributions to the energy variation of electrons (top row) and ions (bottom row) from the cumulative integral of $\JdotE$, $\PiD$, and $\ptheta$. The individual terms in $\JdotE$ and $\PiD$ are shown in the middle and right columns, respectively.}
\label{fig:PiD}
\end{figure*}

The time-dependent energetics of the system is shown in Figure \ref{fig:dE}, where we report the evolution of total electromagnetic energy $\EEM= (B^2+E^2)/8\pi $, kinetic bulk energy $\Ef=\rho u^2/2$ (where $\rho$ is the mass density), and kinetic thermal energy $\Eth = (m/2) \int (\vecv-\vecu)^2 f d\vecv$ (where $f$ is the particle distribution function), for ions and electrons in the whole domain, normalized to the initial total energy of the system. 
We observe a continuous decrease in $\EEM$ and $\Ef_i$ (which corresponds to a decrease in the AW energy), and a simultaneous increase in $\Eth_i$.

To better characterize this energy transfer, we quantify the energy conversion channels \cite{yang2017}. For a periodic system, the rate of change of $\Ef$ and $\Eth$ for each species $\alpha$ is
\begin{equation}
    \frac{\partial\langle \Ef_\alpha\rangle}{\partial t} = \langle \vecJ_\alpha\cdot\vecE\rangle + \langle \matPi_\alpha : \matD_\alpha\rangle + \langle p_\alpha\theta_\alpha\rangle,
    \label{eq:dEfdt}
\end{equation}
\begin{equation}
    \frac{\partial\langle E_\alpha^\mathrm{th}\rangle}{\partial t} = - \langle \mathbf{\Pi}_\alpha : \textbf{D}_\alpha\rangle - \langle p_\alpha\theta_\alpha\rangle,
    \label{eq:dEthdt}
\end{equation}
where $\langle ...\rangle$ denotes spatial average. From Maxwell's equations, the rate of change of the average electromagnetic energy is
\begin{equation}
    \frac{\partial\langle \EEM\rangle}{\partial t} = -\sum_\alpha\langle \vecJ_\alpha\cdot\vecE\rangle.
    \label{eq:dEemdt}
\end{equation}
Equations \eqref{eq:dEfdt}--\eqref{eq:dEemdt} completely characterize the system's energy flows: the electromagnetic fields exchange energy with the plasma kinetic bulk energy by means of the $\JdotE$ terms; each species can convert their own bulk energy into thermal energy via the $\PiD$ and $\ptheta$ terms, and vice versa. The isotropic term $\ptheta$ (with $p=\mathrm{tr}(\matP)/3$ and $\theta=\nabla\cdot\vecu$) is related to the compression and expansion of fluid elements. The deviatoric term $\PiD$, instead, is given by the traceless pressure tensor $\matPi=\matP-p\textbf{I}$ and the stress tensor $\matD=(\nabla\vecu+(\nabla\vecu)^\top)/2-(\theta/3)\textbf{I}$. There is no direct way for the electromagnetic energy to be converted into particle heating, nor can a particle species exchange energy directly with another species, due to the absence of collisions.

\begin{figure*}
\centering
\includegraphics[width=1\textwidth, trim={0mm 75mm 0mm 0mm}, clip]{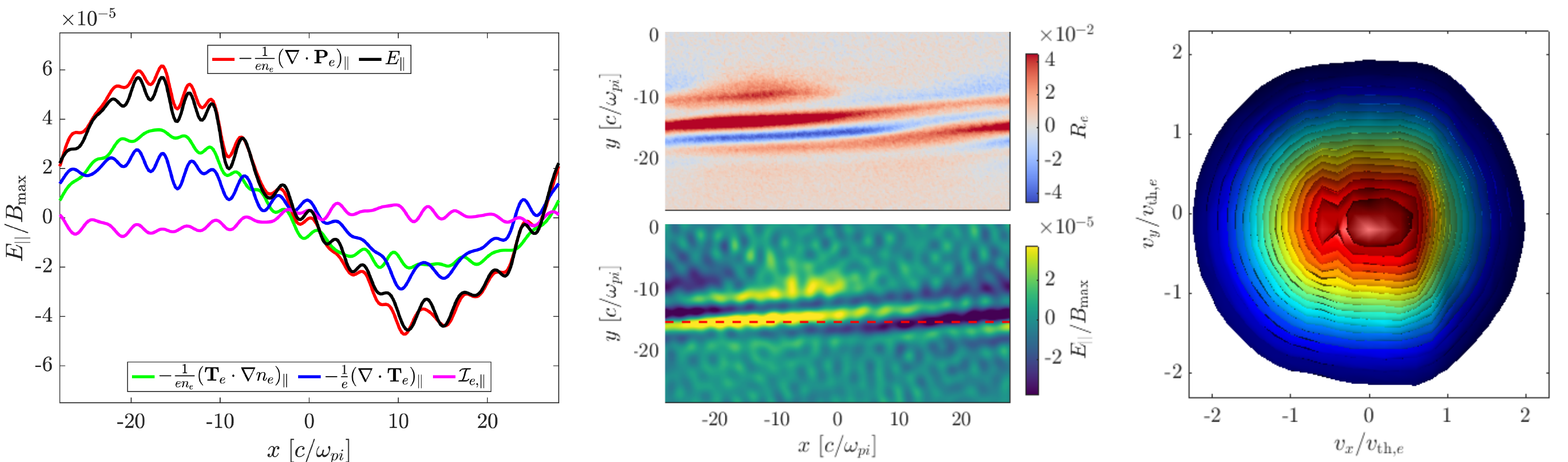}
\caption{Center: Spatial distribution of $\Epar$ and $R_e=1-T_{e,\perp}/T_{e,\|}$ at $t=150\omega_{C,i}^{-1}$ in the lower half of the simulation domain. Left: contribution of the terms in Ohm's law \eqref{eq:ohmslaw} to $\Epar$ along a cut through $y=-15c/\omega_{pi}$; Right: electron VDF measured inside an $R_e>0$ region at $t=150\omega_{C,i}^{-1}$.}
\label{fig:Epar}
\end{figure*}

Figure \ref{fig:PiD} (left column) shows the cumulative time integrals of $\JdotE,\; \PiD$, and $\ptheta$, for electrons (top row) and ions (bottom row). We observe a positive integrated electron $\JedotE$ at almost all times, and a roughly equal and opposite contribution from $\PieDe$. The contribution of $\pthetae$ produces no significant energy gain or loss. Equations \eqref{eq:dEfdt}--\eqref{eq:dEemdt} indicate that $\JdotE>0$ acts as a source for the bulk energy and removes energy from the electromagnetic fields; while $\PiD<0$ removes bulk energy in favor of an increase in thermal energy. 
Therefore, the combined action of the electron $\JdotE$ and $\PiD$ is such that electromagnetic energy is continuously converted into electron internal energy. As a result, a net increase in electron thermal energy is observed.

The ion energetics shows a significantly larger contribution of $\PiiDi$ to the heating, which is not counterbalanced by the action of $\JidotE$ and $\pthetai$. The removal of bulk energy via $\PiD$ is accompanied by a roughly equal increase in thermal energy, with no significant decrease in electromagnetic energy. Since the electromagnetic field cannot be the source of the observed ion heating, we conclude that such heating directly derives from the initial ion bulk motion. Note that the electron bulk energy stored into the initial AW is negligible (since $m_e\ll m_i$), and therefore electrons cannot experience a similar process of bulk-to-thermal energy conversion. Since AWs carry equal amounts of magnetic and kinetic energy, the result presented above indicates that, during the AW-to-KAW conversion, half of the wave energy (i.e.\ the kinetic part) can only be accessed by ions, while the other half (i.e.\ the magnetic part) can in principle be accessed by both species. This is the main cause of the differential heating observed here for ions and electrons, and the reason for the excess in ion-to-electron heating.

In Figure \ref{fig:PiD} we also show the individual terms in the sum composing the integrated $\JdotE=\sum_j J_j E_j$ (middle column) and $\PiD=\sum_{jk} \Pi_{jk}D_{jk}$ (right column). For both particle species the energization due to $\JdotE$ occurs along $x$ (approximately the direction of $\vecB$), which implies the presence of a parallel electric field $\Epar\simeq E_x$. The interaction of ions with $\Epar$ in this specific configuration has been observed in previous works \cite{vasconez2015kinetic}; here, we show that $J_{i,\|}\Epar>0$ is balanced by a $J_{i,\perp}E_{\perp}<0$, resulting in a negligible net $\JidotE$. Instead, the $J_{\|}\Epar$ term is dominant for the electrons, and a net transfer of electromagnetic energy to electron kinetic energy is observed. 
Concerning the effect of $\PiD$, the dominant contributions for ions are both diagonal $\Pi_{i,xx}D_{i,xx}$ and off-diagonal terms $\Pi_{i,xz}D_{i,xz}$, while for electrons the diagonal, field-aligned term $\Pi_{e,xx}D_{e,xx}$ is by far dominant over off-diagonal terms. Considering that such off-diagonal terms stem from agyrotropy in the particle VDFs \cite{delsartopegoraro2018}, this suggests that finite Larmor radius (FLR) effects concur in the conversion of ion kinetic energy into thermal energy, while electrons remain gyrotropic.

The presence of $\Epar$ further supports the interpretation of the observed fluctuations as KAWs, which typically carry a parallel electric field component \cite{hollweg1999kinetic}. The origin of this $\Epar$ can be found in Ohm's law: the parallel part of \eqref{eq:ohmslaw} is
\begin{equation}
 \Epar = -\frac{1}{e n_e}(\nabla\cdot\matP_e)_{\|}+\mathcal{I}_{e,\|},
\end{equation}
and we can expand $\nabla\cdot\matP_e=n_e\nabla\cdot\matT_e+\matT_e\cdot\nabla n_e$, which indicates that a $\Epar$ can arise from both density and temperature gradients, as we observe in our simulation.

During the initial stages of the system's evolution, the generation of $\Epar$ is primarily driven by density gradients caused by phase mixing. As electrons are accelerated by $\Epar$ and the temperature increases, anisotropies in $\matT_e$ also begin contributing to the creation of parallel electric field (Figure \ref{fig:Epar}, left panel). The process then becomes (temporarily) self-fueling: parallel electron acceleration results in parallel current (hence $\JdotE>0$) and is accompanied by heating via $\PiD<0$. This increases the parallel temperature, creates temperature gradients that further contribute to $\Epar$, and so on. As time passes, the initial large-scale AW loses energy and the driving of density fluctuations inside the shear slows down, eventually halting the energy exchange between electrons and fields.

Figure \ref{fig:Epar} also shows the spatial distribution of electron temperature anisotropies $R_e= 1-T_{e,\perp}/T_{e,\|}$ at $t=150\omega_{C,i}^{-1}$. Fluctuations with $R_e>0$ are clearly visible inside the shear. Accordingly, inside these KAW structures we find that the electron VDFs consistently present parallel temperature anisotropy. At subsequent times, the VDFs show a core-beam structure (Figure \ref{fig:Epar}, right panel) that continuously oscillates around $v_x=0$. This oscillation suggests that $\Epar$ of opposite signs are pushing electrons along the $\pm x$ directions equally. We find that the beam consistently develops around the local KAW phase speed, coherently with the development of a beam observed in the ion VDF around the same speed \cite{vasconez2015kinetic}. We note that $R_e<0$ regions also arise inside the shear due to the transition from AW to KAWs via phase mixing. However, these regions rapidly disappear as they drift away from the shear, while $R_e>0$ regions survive until the end of the run.

{\it Discussion and conclusions.} We have presented the first fully kinetic simulation of KAW generation from Alfv\'enic phase mixing in magnetic shears.
The transition from an initial AW to KAWs in an inhomogeneous background yields a net increase in the ion and electron internal energy. The accessible portion of magnetic and kinetic energy in the initial wave is different for ions and electrons, resulting in a net differential heating for the parameters employed in our simulation.

From the kinetic point of view, the measured particle heating is partially promoted by the fact that both ions and electrons resonate with the $\Epar$ carried by KAWs, in a process akin to Landau damping. This electric field is sustained by both density and electron temperature gradients, suggesting that the isothermal closure for electrons employed in previous studies \cite{vasconez2015kinetic,pucci2016alfven,valentini2017transition,maiorano2020kinetic} may be insufficient in describing the electron physics. In this Letter we have explored the physics of electrons in such a scenario for the first time, showing the development of non-Maxwellian features in the electron VDFs. We also observed that ions experience FLR effects, which are absent for electrons. FLR contributions are hinted at in our $\PiD$ analysis, which suggests that these effects promote and participate in the conversion of ion bulk energy into internal energy.

The phase mixing process transforms initially parallel fluctuations into increasingly smaller perpendicular ones. Although such a spectral energy transfer may resemble the phenomenology of turbulence, phase mixing remains a very distinct process. In our simulation, we observe an accumulation of energy in the spectral region where $k_y r_{L,i}=1$, indicating a continuous formation of a narrow-band population of KAWs (see Supplemental Material). This energy cascade does not proceed further towards larger $k_y$. When KAW-like fluctuations form, they immediately start drifting out of the shear, causing the phase mixing to halt. This is different from the phenomenology of turbulence, in which energy cascades towards scales smaller than $r_{L,i}$ (see e.g.\ \citep{kiyani2015dissipation}).

Theoretical studies of Alfv\'enic turbulence have reported a preferential electron heating for systems with $\beta\leq 1$ and a preferential ion heating for $\beta > 1$ \cite{quataert1998particle,howes2011gyrokinetic,told2015multiscale,klein2017diagnosing,parashar2018dependence,kawazura2019thermal}. Our results suggest a possibly different behavior of linear phase mixing with respect to non-linear turbulence in partitioning energy between ions and electrons.
In the Supplemental Material, we explain why this discrepancy goes beyond linear theory of KAW Landau damping.
However, we emphasize that the energy argument presented above only prescribes the lower and upper limits of electron and ion energization. Due to the resonant nature of the processes at play, the actual gain in energy experienced by both species will likely depend on $\beta$ of each species~\cite{schekochihin2019constraints}, on the mass ratio $m_i/m_e$, and on the temperature ratio $T_i/T_e$. Such a parametric study is left for future work.

Phase mixing is an entirely generic process that solely requires the interaction of large-scale fluctuations with gradients in background velocity, density, and/or magnetic field strength to take place. As such, it is entirely plausible that it may play a relevant role in a number of astrophysical systems.
For example,  
it may constitute the generation mechanism for the observed signatures of parallel electron heating and acceleration during KAW generation from large-scale AWs~\cite{wygant2002evidence} and for the electron core-beam structures that have recently been detected within more accurate observations of the Earth's magnetosphere~\cite{gershman2017wave}.
It is also expected that kinetic-scale phase mixing plays a more dominant role in several other astrophysical scenarios where direct measurements are not available, e.g.\ the solar corona \cite{heyvaerts1983coronal}, or the magnetosphere of compact objects \cite{chen2020}. This lack of direct measurements, along with a lack of theoretical work on phase-mixing at kinetic scale, may have contributed to underestimating the importance of this process.
Finally, it is worth recalling that AW phase mixing has long been considered a viable route for the release of wave energy in fusion plasmas \cite{chen1974plasma,hasegawa1974plasma}, which makes the results presented in this Letter potentially interesting beyond plasma astrophysics.

In conclusion, despite its simplicity, the model presented in this Letter constitutes a paradigm for the fate of Alfv\'en waves in magnetic shears at kinetic scales, and is in principle applicable to a wide range of astrophysical and laboratory plasmas where Alfv\'en waves and magnetic shears are present.

\begin{acknowledgements}
The computational resources and services used in this work were partially provided by the VSC (Flemish Supercomputer Center), funded by the Research Foundation -- Flanders (FWO) and the Flemish Government -- department EWI. We acknowledge PRACE for awarding us access to SuperMUC-NG at GCS@LRZ, Germany. F.B.\ is supported by a PostDoctoral Fellowship 12ZW220N from Research Foundation -- Flanders (FWO).
F.P.\ is supported by the PostDoctoral Fellowship 12X0319N and the Research Grant 1507820N from Research Foundation -- Flanders (FWO). This work received support from the European Union’s Horizon 2020 research and innovation programme under grant agreement No.\ 776262 (AIDA, \texttt{www.aida-space.eu}).

F.B. and F.P. contributed equally to this work.

\end{acknowledgements}

%

\end{document}